\documentclass[aps,twocolumn,amsfonts,amssymb,prl,superscriptaddress,10pt,lengthcheck
]{revtex4-2}

\usepackage{graphicx}
\usepackage{xcolor}
\usepackage{amsmath}
\usepackage{hyperref}
\usepackage{orcidlink}
\hypersetup{colorlinks=true,citecolor=blue,linkcolor=blue,urlcolor=blue}

\allowdisplaybreaks 

\begin{document}

\title{Thermopower overshoot from collinear scattering in two-dimensional Fermi liquids}

\author{Johannes Hofmann\,\orcidlink{0000-0002-0667-2452}}
\email{johannes.hofmann@physics.gu.se}
\affiliation{Department of Physics, Gothenburg University, 41296 Gothenburg, Sweden}
\affiliation{Nordita, Stockholm University and KTH Royal Institute of Technology, 10691 Stockholm, Sweden}

\date{\today}

\begin{abstract}
Electron transport is typically described in terms of relaxation times, which measure how quickly Fermi-surface deformations decay. We show here that the thermopower of a two-dimensional Fermi liquid instead probes the structure of these deformations in the form of their local temperature change, which we call the thermal dressing. This structure leads to an overshoot of the thermopower over its hydrodynamic value $- s/(ne)$ with increasing wave vector, before it falls to half that value in the ballistic limit. At low temperatures, the excess thermopower does not have the linear-in-temperature scaling expected of a Fermi liquid, but decreases only logarithmically as $1/\ln(T_F/T)$ at optimal wave vector. The overshoot is caused by collinear scattering, which redistributes energy over the Fermi surface and induces the thermal dressing, the strength of which grows with the angular harmonic index $m$ of the deformation. In contrast to tomographic effects in the charge conductivity, this overshoot does not rely on an odd-even effect in quasiparticle lifetimes and defines a distinct thermographic transport regime. The thermopower overshoot survives realistic disorder and should be detectable in a thermocouple with high-mobility samples. 
\end{abstract}

\maketitle

The thermopower ${\cal S}$ measures the entropy carried in a conductor per unit charge~\cite{callen48,behnia15}. In a hydrodynamic metal, in which momentum-conserving binary collisions between carriers dominate, the thermopower is equal to ${\cal S}_{\rm hyd} = - s/(ne)$~\cite{hartnoll07,lucas18,ghahari16,xie16}, independent of sample dimensions, and in the ballistic limit, it is halved. One could then expect that the crossover between these regimes is featureless. In this Letter, we show that this is not the case in two-dimensional Fermi liquids: At finite wave vectors,  as realized in a narrow channel, the thermopower overshoots the hydrodynamic value by a  factor that grows as $T_F/[T \ln(T_F/T)]$ ($T_F$ is the Fermi temperature, and we set $k_B=1$). 

The cause of the thermopower overshoot is near-collinear scattering, which transfers energy between different points on the Fermi surface and induces a local temperature change in the quasiparticle distribution
\begin{align}
        \delta f_m({\bf k}) &\sim \biggl(- \frac{\partial f_0}{\partial \varepsilon}\biggr) \bigl(\delta \mu_m + \delta T_m 
        w\bigr) e^{im\theta} , \label{eq:odddeformation}
\end{align}
where $\theta$ is the momentum angle on the Fermi surface, $m$ is an angular harmonic index, $w=(\varepsilon-\mu)/T$ with $\varepsilon$ the quasiparticle energy and $\mu$ the chemical potential, $\delta \mu_m$ is a rigid local chemical potential change, and $\delta T_m \propto \delta \mu_m$ is the local temperature change, which we call the thermal dressing. The strength of the thermal dressing is $|\delta T_m/\delta \mu_m| \sim m^2 T/T_F$ [Eq.~\eqref{eq:deltaTdeltamu}]~\cite{boeve26}, which for $m>1$ exceeds that of hydrodynamic flow. At finite wave vector~$q$, which couples the current to higher angular harmonics, this dressing sets the thermopower. The thermopower is thus sensitive to the structure of these modes (mathematically, the eigenvectors of the collision integral) rather than only to their relaxation times (its eigenvalues). 

Near-collinear scattering has so far been discussed mainly in the context of electron relaxation. In two dimensions, energy and momentum conservation no longer act as separate constraints for approximately parallel or antiparallel  scattering momenta, which increases the phase space for collisions~\cite{laikhtman92}. This leads to a characteristic logarithmic enhancement of relaxation rates, the best-known example being the $T^2\ln(T_F/T)$ scaling of the single-particle decay rate~\cite{hodges71,bloom75,giuliani82,zheng96}. Recent work has focused on the relaxation of angular harmonics $e^{im\theta}$ of Fermi-surface deformations, which is marked by an odd-even effect. Here, head-on processes, which affect opposite points on the Fermi surface equally, relax even-parity Fermi-surface deformations (even $m$) but leave the odd-parity components (odd $m$) unchanged. The latter decay by finite-temperature corrections to this kinematics, with rates that are suppressed relative to even rates by $(m^2 T/T_F)^2$~\cite{ledwith19,hofmann23,nilsson25}. This odd-even effect in quasiparticle lifetimes underlies the proposed tomographic transport regime that mixes the dynamics of short-lived even-parity modes with nearly collisionless odd-parity modes~\cite{gurzhi82,gurzhi87,gurzhi93,gurzhi95c,ledwith19b,hofmann22a,kryhin23}. Signatures include nonlocal transport in confined geometries~\cite{ledwith19b,kryhin25,alvarez25,starkov26a,benshachar25c,farrell26,benshachar26}, magnetotransport~\cite{rostami25,maki26,starkov26b,benshachar26b}, and dynamical probes~\cite{moiseenko25,thuillier26}. These analyses assign a single relaxation rate to each angular harmonic (generalizing the relaxation-time approximation~\cite{callaway59,dejong95}), and consider charge transport, which is insensitive to the structure of the deformation at leading order. We show below that the thermopower overshoot does not rely on anomalously long lifetimes of odd modes, and we call the regime where it occurs the thermographic regime, to distinguish it from tomographic transport.

The link between the thermal dressing of a mode and the thermopower follows from the fact (by Onsager reciprocity) that the thermopower is equal to the ratio of the entropy current to the charge current ${\cal S} = j_s/j$. In a Fermi liquid, $j_s = \langle w v\rangle$ and $j=\langle -e v \rangle$, where $\langle \ldots \rangle$ is the average over the deviation $\delta f$ from equilibrium, $-e$ is the electron charge, and $v \simeq v_F [1+wT/(2T_F)]$ is the particle velocity to leading order in $T/T_F$. For a moving local Fermi-Dirac distribution of the form~\eqref{eq:odddeformation}, a direct expansion gives
\begin{align}
{\cal S} = \frac{1}{2} \bigl(1 + \vartheta\bigr) {\cal S}_{\rm hyd} , \label{eq:thermopowerovershoot}
\end{align}
where ${\cal S}_{\rm hyd} = - (\pi^2/3) T/(eT_F)$ is the value of $-s/(ne)$ for a parabolic band, and we define the tilt $\vartheta = (2T_F/T) \delta T/\delta \mu$ as a dimensionless measure of the thermal dressing. A Galilean boost, $f \simeq f_0 + \hbar {\bf k} \cdot {\bf u} (-\partial f_0/\partial \varepsilon)$, has $\vartheta=1$, corresponding to the hydrodynamic thermopower, and a rigid deformation has $\vartheta=0$ and hence ${\cal S}/{\cal S}_{\rm hyd} = 1/2$. While a current is described by the $m=1$ harmonic, at finite wave vectors it couples to higher harmonics and can in this way inherit the thermal dressing of higher modes. Such a dressed mode with $\vartheta = m^2 \xi_m$ has ${\cal S}/{\cal S}_{\rm hyd} = (1+m^2 \xi_m)/2$, which exceeds the hydrodynamic value for $m^2\xi_m > 1$  [where $0 \leq \xi_m \leq 1$  is the relative dressing of Ref.~\cite{boeve26}, which accounts for a reduction of the thermal dressing due to noncollinear large-angle scattering, 
cf.~Eq.~\eqref{eq:xim}].

\begin{figure}
	\centering
    \scalebox{1}{\includegraphics{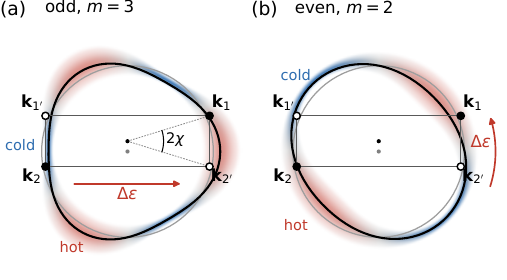}}
    \caption{
    Near-collinear scattering process ${\bf k}_1 + {\bf k}_2 \leftrightharpoons {\bf k}_{1'} + {\bf k}_{2'}$, with filled (open) circles showing initial (final) states, on a background local Fermi-surface deformation (black) with (a) odd parity $m=3$ and (b) even parity $m=2$. The equilibrium Fermi surface (gray) is shown for comparison. 
    Colors indicate the thermal dressing, and red arrows the net energy transfer in the sketched collision process.
    }
    \label{fig:1}
\end{figure}

To illustrate the effect of collinear scattering on quasiparticle relaxation and thermal dressing, consider a near-collinear momentum configuration ${\bf k}_1 + {\bf k}_2 \leftrightharpoons {\bf k}_{1'} + {\bf k}_{2'}$ [Fig.~\ref{fig:1}]. The figure also shows a rigid Fermi-surface deformation (black lines) of the form  $\delta \mu(\theta) = \delta \mu \cos(m(\theta-\theta_0))$ with $m=3$ and $\theta_0=0$ [odd parity, Fig.~\ref{fig:1}(a)] and $m=2$ and $\theta_0=-\pi/4$ [even parity, Fig.~\ref{fig:1}(b)]. By energy and momentum conservation, the endpoints of the four scattering momenta form a rectangle. For nearly head-on collisions with scattering momenta not exactly antiparallel, its center (black dot) is displaced from the center of the Fermi surface (gray dot) by a small amount $k_F d$. The momenta then sit at $\theta_{1,2'}=d\pm\chi$ and $\theta_{2,1'}=\pi-d\pm\chi$, and scattering transfers the energy $\Delta \varepsilon = \varepsilon_1 - \varepsilon_{2'} = 4T_F d \chi \lesssim T$ across the Fermi surface. The bias between the forward and reverse processes is set by the ratio of the local Fermi factors 
\begin{align}
\ln \frac{\Gamma_\rightharpoonup}{\Gamma_\leftharpoondown} &= {\sum_i}^\prime \ln \frac{f_i}{1-f_i} \nonumber \\ 
&\simeq \frac{1}{T} {\sum_i}^\prime [\delta \mu(\theta_i) + w_{i} \delta T(\theta_{i})] \equiv r_\mu + r_T ,
\end{align}
where $\sum^\prime$ runs over the initial minus final states and we used $f/(1-f)=e^{-[\varepsilon-\mu(\theta)]/T(\theta)}$ with local changes $\delta \mu(\theta) = \delta \mu_m e^{im\theta}$ and $\delta T(\theta) = \delta T_m e^{im\theta}$ of the chemical potential and temperature.

For an odd-parity deformation, $\delta \mu(\theta + \pi) = - \delta \mu(\theta)$ and likewise $\delta T(\theta + \pi) = - \delta T(\theta)$. A direct Taylor expansion then gives $r_\mu = 4 \chi d \delta \mu''(0)/T$, i.e., the gradient term cancels between the opposite patches and the bias is proportional to the curvature of the Fermi-surface deformation. Since $\delta \mu''(0)<0$ for $\delta \mu>0$, this biases the scattering in Fig.~\ref{fig:1}(a) in the reverse direction and leads to a gradual net transfer from regions of excess quasiparticles to regions with a depletion of quasiparticles, so collinear scattering starts to relax the mode. At the same time, the net energy transfer $\Delta \varepsilon$ induces a local temperature imbalance between the right and left patches, with a corresponding thermal bias $r_T = 2 \delta T(0) \Delta \varepsilon/T^2 \sim \chi d$. Since both biases scale as $\chi d$, they balance for $\delta T(0) = - T\delta \mu''(0)/(2T_F)$, which gives 
\begin{align}
    \frac{\delta T_m}{\delta \mu_m} &= \frac{m^2}{2} \frac{T}{T_F} . \label{eq:deltaTdeltamu}
\end{align}
Patches with an excess of quasiparticles are hotter and those with a depletion are colder [red and blue in Fig.~\ref{fig:1}(a)]. Intuitively, the heating on hot patches leads to an increased occupation of higher-energy states, which now biases the collinear scattering in the forward direction. This implies a detailed balance between forward and backward scattering for every near-collinear collision, such that collinear scattering does not relax the dressed odd-parity deformation. Large-angle scattering then provides the only remaining relaxation path. For $m=1$ (the Galilean boost), the thermal dressing also protects against large-angle scattering, and the mode has infinite lifetime, as required by momentum conservation. Since the origin of thermal dressing is kinematic, we use a constant interaction potential in the following, for which a low-temperature expansion of the collision integral to leading order in $T/T_F$ gives~\cite{boeve26}
\begin{align}
\xi_m = 1 - \frac{2}{\bar{\Lambda}_T} \Bigl(2 o_m - 1 - \frac{1}{m}\Bigr) , \label{eq:xim}
\end{align}
where $o_m = 
\sum_{{\rm odd} \, j \leq m} 1/j \simeq (\ln2m)/2 \allowbreak + \gamma_E/2 + {\it O}(1/m)$ with $\gamma_E$ the Euler constant and $\bar{\Lambda}_T = \ln(e^hT_F/T)$ with $h = -0.05615$. At $m=1$, we have $o_1=1$ and hence $\xi_1=1$. With increasing $m$, $\xi_m$ gradually decreases and reaches zero for $m^2 T/T_F = {\it O}(1)$, which marks the boundary where collinear scattering no longer dominates over large-angle relaxation.

For an even-parity deformation [Fig.~\ref{fig:1}(b)], $\delta \mu(\theta + \pi) = \delta\mu(\theta)$ and likewise $\delta T(\theta + \pi) = \delta T(\theta)$. Here, $r_\mu = 4 \chi \delta \mu'(0)/T$ depends on the gradient of the Fermi-surface deformation, and the transferred energy vanishes on average, without any net transfer between Fermi-surface patches. The thermal dressing is then set by a competition between its generation by collinear scattering and energy relaxation rather than by detailed balance. The kinetic-theory calculation gives $\delta T/\delta \mu \sim - m^2 T/T_F$ [see End Matter and App.~E of Ref.~\cite{boeve26}], which has the opposite sign to the odd-mode case, i.e., Fermi-surface patches with an excess of quasiparticles are colder [Fig.~\ref{fig:1}(b)]. 

To see how the thermal dressing manifests in transport at finite wave vector, we solve the linearized kinetic equation. For a general deviation $\delta f = (-T\partial f_0/\partial \varepsilon) h$, it reads
\begin{align}
&\frac{1}{T} \biggl[
\frac{(\varepsilon - \mu)}{ T} {\bf v} \cdot {\boldsymbol \nabla}_\perp T 
+
{\bf v} \cdot e {\bf E}_\perp 
\biggr] + i {\bf v} \cdot {\bf q} h = - {\cal L}[h] . \label{eq:kinetic}
\end{align}
We include transverse plane-wave perturbations in the temperature gradient and the electric field $\nabla_\perp T e^{+ i {\bf q} \cdot {\bf r}}$ and $E_\perp e^{+ i {\bf q} \cdot {\bf r}}$, with a wave vector ${\bf q} = (q,0)$ along the $x$ direction and $\nabla_\perp T$ and $E_\perp$ along the $y$ direction. Such a setup provides an effective description of flow in a finite channel of width $W=\pi/q$~\cite{ledwith19b,rostami25,zeng24,benshachar26,thuillier26}. 

\begin{figure}
	\centering
    \scalebox{1}{\includegraphics{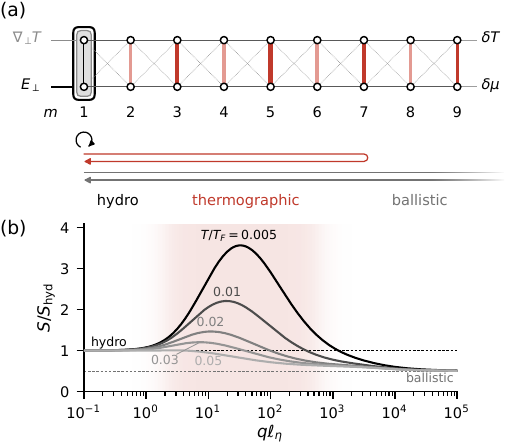}}
    \caption{(a) Tight-binding representation of the kinetic equation~\eqref{eq:projectedkinetic}. Each rung represents an angular harmonic $m$, and the lower (upper) leg represents the rigid $\delta \mu$ (temperature $\delta T$) component. Adjacent harmonics are coupled by the advective term $(v_Fq/2) {\cal V}$ (horizontal lines and gray crosses), and electric (thermal) probes couple to the  charge (heat) current at $m=1$ (boxes). Red rungs indicate the thermal dressing, with the thickness indicating its strength, which is largest for $m \lesssim \sqrt{T_F/T}$. (b) Thermopower as a function of wave vector $q\ell_\eta$ for five temperatures $T/T_F=0.005,0.01,0.02,0.03$, and $0.05$. 
    }
    \label{fig:2}
\end{figure}

The collision integral is positive semidefinite with respect to the inner product $\langle h | h' \rangle = (N_0 T)^{-1} \int_{\bf k} f_0(1-f_0) h^* h'$, where $N_0 = m^*/(2\pi\hbar^2)$ is the density of states per spin. The first orthonormal basis vectors are $u_0 = 1$ and $u_1 = \sqrt{3} w/\pi$, which describe a rigid shift and a local temperature change. Truncating the radial basis after these two elements, we obtain an effective description in terms of the two-component vector ${\bf h}_m = (h_\mu, h_T)^T$. The first  component is the local change $h_\mu = \delta \mu(\theta)/T$, and the second $h_T = (\pi/\sqrt{3}) \delta T(\theta)/T$. 
The projected kinetic equation is
\begin{align}
    &i \frac{v_F q}{2} {\cal V} \bigl[{\bf h}_{m+1} + {\bf h}_{m-1}\bigr] + L_m {\bf h}_m \nonumber \\
    &\quad = - \delta_{m,1} \frac{v_F}{T} (e E_\perp {\bf c} + \nabla_\perp T {\bf s}) , \label{eq:projectedkinetic}
\end{align}
where $L_m$ is the $2\times2$ collision matrix in angular subspace $m$, ${\cal V} = I + \kappa \sigma_x$ represents  advection (with $\sigma_x$ a Pauli matrix and $\kappa = \pi T/(2\sqrt{3}T_F)$), and the source vectors  ${\bf c} = {\cal V} {\bf e}_0 = (1,\kappa)^T$ and ${\bf s} = (\pi/\sqrt{3}) {\cal V} {\bf e}_1 = (\pi/\sqrt{3}) (\kappa,1)^T$ induce a  charge current and a heat current, respectively. Equation~\eqref{eq:projectedkinetic} takes the form of a tight-binding ladder [Fig.~\ref{fig:2}(a)], with one rung for each angular harmonic and advection as hopping between adjacent rungs, and sources act at $m=1$ [boxes in Fig.~\ref{fig:2}(a)]. Since transverse sources only excite $\sin (m\theta)$ harmonics with $m\geq 1$, the ladder starts at $m=1$. 

In linear response, the charge current is ${\bf j} = 
- 2 e N_0 T \langle {\bf v} | h \rangle $. For ${\bf q}$ along the $x$ axis, only the $y$ component is nonzero, $j_y = \sigma E_\perp - \alpha \nabla_\perp T$, with $\sigma \sim e^2 {\bf c}^T {\cal G}_1 {\bf c}$ the charge conductivity and $\alpha \sim - e {\bf c}^T {\cal G}_1 {\bf s}$ the thermoelectric coefficient (up to common prefactors). The propagator ${\cal G}_m$ satisfies the Dyson equation
\begin{align}
{\cal G}_m &= \Bigl[L_m + \Bigl(\frac{v_F q}{2}\Bigr)^2 {\cal V} {\cal G}_{m+1} {\cal V}\Bigr]^{-1} , \label{eq:recursion}
\end{align}
which is the matrix equivalent~\cite{risken96} of a continued-fraction expansion for a single-orbital chain~\cite{hofmann22a,nazaryan24}. ${\cal G}_m$ describes excursions from site $m$ to higher sites and back. Its inverse consists of the on-site decay $L_m$ and the self-energy term $-(v_F q/2)^2 {\cal V} {\cal G}_{m+1} {\cal V}$, which describes irreducible excursions, i.e., those that leave site $m$ and return exactly once. At open circuit $j_y=0$, the thermopower (or Seebeck coefficient) is the ratio of the electric field $E_\perp$ that builds up to the applied thermal force $\nabla_\perp T$,
\begin{align}
{\cal S} &= \frac{\alpha}{\sigma} = - \frac{1}{e} \frac{{\bf c}^T {\cal G}_1 {\bf s}}{{\bf c}^T {\cal G}_1 {\bf c}} . \label{eq:thermopower}
\end{align}
Since ${\cal G}_1$ is symmetric, this expression measures the overlap of a propagated charge boost ${\cal G}_1 {\bf c}$ with the heat current normalized to the overlap with the original charge boost. Since $c_0 = {\it O}(1)$ and $c_1 = {\it O}(T/T_F)$ and mixing occurs by the off-diagonal elements of ${\cal V}$ and $L$, the same hierarchy will hold for the return vector ${\cal G}_1 {\bf c}$. Writing ${\cal G}_1 {\bf c} \sim \delta \mu (1,\kappa \vartheta)$, we again obtain Eq.~\eqref{eq:thermopowerovershoot}. 

Figure~\ref{fig:2}(b) shows the thermopower computed using the two-moment recursion~\eqref{eq:recursion} with analytical collision-matrix elements for a constant interaction potential [see End Matter]. Wave vectors are stated in units of $1/\ell_\eta = \gamma_\eta/v_F$, which is set by the $m=2$ shear relaxation rate. At small~$q$, the thermopower is fixed at the hydrodynamic value ${\cal S}/{\cal S}_{\rm hyd} = 1$. In the representation of Fig.~\ref{fig:2}(a), excursions from the initial site $m=1$ are suppressed as ${\it O}(q^2)$ and thus the return vector equals the charge boost. For $q\ell_\eta \gtrsim 1$, the initial charge boost explores an increasing number of sites before returning. Modes that remain in a given angular harmonic long enough before being advected further or decaying pick up a thermal dressing by collinear scattering. If this dressing propagates back to site $m=1$, this gives rise to an overshoot in the thermopower above the hydrodynamic value, which grows with decreasing temperature. 
Even modes enhance the thermopower despite their negative dressing $\delta T \sim - m^2 (T/T_F) \delta \mu$, since an excursion to site $m$ and back to $m=1$ comes with an advective hopping factor $(iv_Fq/2)^{2(m-1)} \sim (-1)^{m-1}$, which alternates between even and odd sites. 
For $q\ell_\eta \gg 1$, the initial charge boost passes through thermally dressed sites quickly and returns via large excursions without a thermal dressing. From Eq.~\eqref{eq:recursion}, we have ${\cal G}_1 = (2/v_F q) {\cal V}^{-1}$ for $L_m \to 0$, such that the initial boost returns as a rigid deformation ${\cal G}_1 {\bf c} \sim {\bf e}_0$, and ${\cal S}/{\cal S}_{\rm hyd}=1/2$.

\begin{figure}
	\centering
    \scalebox{1}{\includegraphics{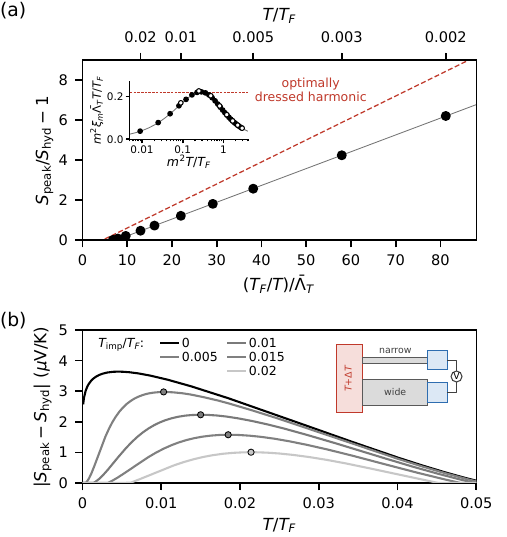}}
    \caption{(a) Thermopower peak relative to the hydrodynamic value as a function of the scaling variable $T_F/(T\bar{\Lambda}_T)$. 
    Inset: Scaling form of the tilt $m^2\xi_m$ as a function of $m^2T/T_F$. Red dashed lines: maximal tilt (inset) and optimal-harmonic estimate (main panel). (b) Excess of the peak thermopower over the hydrodynamic value as a function of temperature for a thermocouple with wide and narrow legs (inset). Gray lines show the excess in the presence of short-range disorder, dots indicate the maxima.}
    \label{fig:3}
\end{figure}

The position of the peak is set by the wave vector at which the response reaches strongly dressed harmonics. From Eq.~\eqref{eq:xim}, the number of such modes is $m_\star = {\it O}(\sqrt{T_F/T})$. Within the two-moment model, the temperature component on the upper leg of Fig.~\ref{fig:2}(a) must return to site $m=1$ before it decays at the collinear rate $\gamma_c$ [$G_m$ in Eq.~\eqref{eq:collisionmatrix}], which exceeds $\gamma_\eta$ by a factor $\sim \bar{\Lambda}_T$. The number of sites covered on the upper leg is given by the ratio $(v_Fq/2)/\gamma_c$ of the hopping rate to the damping, hence, $v_Fq/2 \simeq \gamma_c m_\star$ gives $q\ell_\eta \simeq \bar{\Lambda}_T \sqrt{T_F/T}$. The position of this peak is much larger than the wave vector that sets the onset of tomographic effects in the transverse conductivity $q \simeq 1/\ell_{\rm tom} = \sqrt{\gamma_\eta\gamma_3}/v_F \ll 1/\ell_\eta$, which is set by the relaxation rate $\gamma_3$ of the longest-lived odd mode. Indeed, removing the long odd-mode lifetimes in our description does not remove the thermopower overshoot, whereas removing the thermal dressing eliminates it [cf.~Fig.~\ref{fig:4} in the End Matter]. 

For the height of the peak, Eq.~\eqref{eq:xim} implies that the optimally dressed harmonic, which maximizes the tilt $m^2 \xi_m$, has $m^2 \xi_m \sim T_F/(T\bar{\Lambda}_T)$, which sets ${\cal S}/{\cal S}_{\rm hyd}-1 \sim T_F/(T\bar{\Lambda}_T)$. This scaling is demonstrated in Fig.~\ref{fig:3}(a), which shows the peak height for different temperatures as a function of the scaling variable $T_F/(T\bar{\Lambda}_T)$. A fit [black line] gives ${\cal S}/{\cal S}_{\rm hyd}-1\simeq 0.08 \, T_F/(T\bar{\Lambda}_T) -1/2$ for $T \ll T_F$. The inset shows $m^2 \xi_m/[T_F/(T\bar{\Lambda}_T)]$ as a function of $m^2 T/T_F$ for two different temperatures $T/T_F=10^{-2}$ [open circles] and $T/T_F=10^{-3}$ [filled circles]. Its maximum [red dashed line in inset] gives $\vartheta_{\rm max}/2 = 0.11 T_F/(T\bar{\Lambda}_T)$ for the maximal tilt of a mode, which by Eq.~\eqref{eq:thermopowerovershoot} provides an optimal-harmonic estimate for the peak [red dashed line in Fig.~\ref{fig:3}(a)]. The full response, which includes several angular harmonics, reaches about $70\%$ of the optimal slope. For the peak position, we obtain the fit $q_\star \ell_\eta \simeq 0.47 \bar{\Lambda}_T \sqrt{T_F/T}$.

To observe the thermopower overshoot, we propose a thermocouple setup with a wide and a narrow leg kept between the same temperature reservoirs [inset of Fig.~\ref{fig:3}(b)], which measures the difference in the thermovoltages, $\Delta V = \Delta T [{\cal S}(\pi/W_1) - {\cal S}(\pi/W_2)]$. The black line in Fig.~\ref{fig:3}(b) shows as a function of temperature the predicted excess thermopower of a narrow channel of optimal width relative to a wide hydrodynamic channel. In our model, the overshoot becomes appreciable below $T \approx 0.05T_F$, and then decreases only slowly toward lower temperatures, as $1/\ln(T_F/T)$, in contrast to the linear suppression of ${\cal S}_{\rm hyd} \approx - 283 \mu V/K \times T/T_F$ itself~\cite{behnia04}. To estimate the robustness against disorder, we include elastic impurity scattering in the collision matrix by $L_m \to L_m+\gamma_m^{\rm dis} I$. For short-range disorder, $\gamma_m^{\rm dis} \equiv \gamma^{\rm dis}$ is constant, which we parametrize by a disorder temperature $T_{\rm imp}$ defined by $\gamma^{\rm dis} = \gamma_\eta(T_{\rm imp})$ [gray curves in Fig.~\ref{fig:3}(b)]. The excess thermopower is maximal at $T=(1$--$2)T_{\rm imp}$, so that the  experimental window is $T_{\rm imp} \lesssim T \lesssim 0.05 T_F$. Disorder rates that grow with the angular harmonic index $m$, such as for scattering off remote impurities, reduce the excess further since they damp higher harmonics at a rate closer to the quantum rate $1/\tau_{\rm q}$ than to the transport rate $1/\tau_{\rm tr}$ [see End Matter]. These requirements on disorder rates are much looser than those for  tomographic transport, which requires disorder rates below the small odd-mode relaxation rates. 

A related thermocouple setup 
was realized by \citet{chickering09}, who measured the thermovoltage in a GaAs/AlGaAs two-dimensional electron gas using a hot-electron technique that minimizes phonon drag. The experiment varies the electron density rather than the channel width up to $T\approx 2K$ (corresponding to $T/T_F \lesssim 0.03$, the range relevant here). While this experiment is in a diffusive regime, thermometry in hydrodynamic GaAs narrow channels has recently been reported~\cite{pusep26}, which combined with a thermoelectric voltage measurement as in Ref.~\cite{chickering09} would give access to the thermographic regime. Focusing on GaAs and a  density $n=4.2 \cdot 10^{11} cm^{-2}$~\cite{wang22}, and using the estimate $\gamma_\eta \simeq (0.6$--$0.8) T^2/(\hbar T_F)$~\cite{pusep26,wang22}, our fit for $q_\star \ell_\eta$ gives channel widths of $W \simeq (10$--$20)\mu m$, and since the thermopower peak is quite broad, this should not require much fine tuning. Observing the peak requires high-mobility samples: For  short-range disorder with $T_{\rm imp}/T_F = 0.02$ (corresponding to a mobility \mbox{$\mu_e \approx (3.5$--$4.6) \cdot 10^6 cm^2/(Vs)$} in the range of current hydrodynamic GaAs samples~\cite{pusep26}), the excess thermopower is up to \mbox{$1 \mu V/K$} [Fig.~\ref{fig:3}(b)]. Ultrahigh-mobility samples with $\mu_e = 4.25 \cdot 10^7 cm^2/(Vs)$~\cite{wang22,chung22}, corresponding to $T_{\rm imp}/T_F=0.007$, still reach an excess of \mbox{$1$--$1.5 \mu V/K$} near $T\approx 0.02T_F$ for typical ratios $\tau_{\rm tr}/\tau_{\rm q}=10$--$100$~\cite{dassarma85,macleod09} of small-angle scattering [see End Matter, Fig.~\ref{fig:5}]. An additional setup in which our predictions could be tested is ultracold quantum gases~\cite{brantut13,maki25}. These setups are free of disorder and realize the constant interaction discussed here, although current experiments are limited in the reachable $T/T_F$.

In summary, we show that for wave vectors $1/\ell_\eta \lesssim q \lesssim \sqrt{T_F/T} \ln (T_F/T)/\ell_\eta$, the thermoelectric response of a two-dimensional Fermi liquid is set by thermally dressed harmonics, with a thermopower that exceeds the hydrodynamic value by a factor $\sim T_F/[T\ln(T_F/T)]$. This thermographic regime does not require long odd-mode lifetimes and should be within the reach of current high-mobility devices. More generally, the thermopower overshoot is an example of a response that is set by the structure of Fermi-liquid modes and not just their lifetimes.

\begin{acknowledgments}
I thank Nitay Ben-Shachar, Stan Boeve, Ulf Gran, Hadrien Kurkjian, and Carl L\"ov for discussions.
This work is supported by Vetenskapsr\r{a}det (Grant No.~2024-04485), the Olle Engkvist Foundation (Grant No.~233-0339), the Knut and Alice Wallenberg Foundation (Grant No.~KAW 2024.0129), and Nordita. I used Claude (Anthropic, Opus 5.5) to assist with code for the numerical calculations of the kinetic equations, and checked results against independent implementations. I take full responsibility for all results.
\end{acknowledgments}

\bibliography{bib_oddmodes}

\begin{thebibliography}{49}%
\makeatletter
\providecommand \@ifxundefined [1]{%
 \@ifx{#1\undefined}
}%
\providecommand \@ifnum [1]{%
 \ifnum #1\expandafter \@firstoftwo
 \else \expandafter \@secondoftwo
 \fi
}%
\providecommand \@ifx [1]{%
 \ifx #1\expandafter \@firstoftwo
 \else \expandafter \@secondoftwo
 \fi
}%
\providecommand \natexlab [1]{#1}%
\providecommand \enquote  [1]{``#1''}%
\providecommand \bibnamefont  [1]{#1}%
\providecommand \bibfnamefont [1]{#1}%
\providecommand \citenamefont [1]{#1}%
\providecommand \href@noop [0]{\@secondoftwo}%
\providecommand \href [0]{\begingroup \@sanitize@url \@href}%
\providecommand \@href[1]{\@@startlink{#1}\@@href}%
\providecommand \@@href[1]{\endgroup#1\@@endlink}%
\providecommand \@sanitize@url [0]{\catcode `\\12\catcode `\$12\catcode
  `\&12\catcode `\#12\catcode `\^12\catcode `\_12\catcode `\%12\relax}%
\providecommand \@@startlink[1]{}%
\providecommand \@@endlink[0]{}%
\providecommand \url  [0]{\begingroup\@sanitize@url \@url }%
\providecommand \@url [1]{\endgroup\@href {#1}{\urlprefix }}%
\providecommand \urlprefix  [0]{URL }%
\providecommand \Eprint [0]{\href }%
\providecommand \doibase [0]{https://doi.org/}%
\providecommand \selectlanguage [0]{\@gobble}%
\providecommand \bibinfo  [0]{\@secondoftwo}%
\providecommand \bibfield  [0]{\@secondoftwo}%
\providecommand \translation [1]{[#1]}%
\providecommand \BibitemOpen [0]{}%
\providecommand \bibitemStop [0]{}%
\providecommand \bibitemNoStop [0]{.\EOS\space}%
\providecommand \EOS [0]{\spacefactor3000\relax}%
\providecommand \BibitemShut  [1]{\csname bibitem#1\endcsname}%
\let\auto@bib@innerbib\@empty
\bibitem [{\citenamefont {Callen}(1948)}]{callen48}%
  \BibitemOpen
  \bibfield  {author} {\bibinfo {author} {\bibfnamefont {H.~B.}\ \bibnamefont
  {Callen}},\ }\bibfield  {title} {\bibinfo {title} {{The Application of
  Onsager's Reciprocal Relations to Thermoelectric, Thermomagnetic, and
  Galvanomagnetic Effects}},\ }\href {https://doi.org/10.1103/PhysRev.73.1349}
  {\bibfield  {journal} {\bibinfo  {journal} {Phys. Rev.}\ }\textbf {\bibinfo
  {volume} {73}},\ \bibinfo {pages} {1349} (\bibinfo {year}
  {1948})}\BibitemShut {NoStop}%
\bibitem [{\citenamefont {Behnia}(2015)}]{behnia15}%
  \BibitemOpen
  \bibfield  {author} {\bibinfo {author} {\bibfnamefont {K.}~\bibnamefont
  {Behnia}},\ }\href
  {https://doi.org/10.1093/acprof:oso/9780199697663.001.0001} {\emph {\bibinfo
  {title} {{Fundamentals of Thermoelectricity}}}}\ (\bibinfo  {publisher}
  {Oxford University Press},\ \bibinfo {address} {Oxford},\ \bibinfo {year}
  {2015})\BibitemShut {NoStop}%
\bibitem [{\citenamefont {Hartnoll}\ \emph {et~al.}(2007)\citenamefont
  {Hartnoll}, \citenamefont {Kovtun}, \citenamefont {M{\"u}ller},\ and\
  \citenamefont {Sachdev}}]{hartnoll07}%
  \BibitemOpen
  \bibfield  {author} {\bibinfo {author} {\bibfnamefont {S.~A.}\ \bibnamefont
  {Hartnoll}}, \bibinfo {author} {\bibfnamefont {P.~K.}\ \bibnamefont
  {Kovtun}}, \bibinfo {author} {\bibfnamefont {M.}~\bibnamefont {M{\"u}ller}},\
  and\ \bibinfo {author} {\bibfnamefont {S.}~\bibnamefont {Sachdev}},\
  }\bibfield  {title} {\bibinfo {title} {{Theory of the Nernst effect near
  quantum phase transitions in condensed matter and in dyonic black holes}},\
  }\href {https://doi.org/10.1103/PhysRevB.76.144502} {\bibfield  {journal}
  {\bibinfo  {journal} {Phys. Rev. B}\ }\textbf {\bibinfo {volume} {76}},\
  \bibinfo {pages} {144502} (\bibinfo {year} {2007})}\BibitemShut {NoStop}%
\bibitem [{\citenamefont {Lucas}\ and\ \citenamefont
  {Das~Sarma}(2018)}]{lucas18}%
  \BibitemOpen
  \bibfield  {author} {\bibinfo {author} {\bibfnamefont {A.}~\bibnamefont
  {Lucas}}\ and\ \bibinfo {author} {\bibfnamefont {S.}~\bibnamefont
  {Das~Sarma}},\ }\bibfield  {title} {\bibinfo {title} {{Electronic
  hydrodynamics and the breakdown of the Wiedemann-Franz and Mott laws in
  interacting metals}},\ }\href {https://doi.org/10.1103/PhysRevB.97.245128}
  {\bibfield  {journal} {\bibinfo  {journal} {Phys. Rev. B}\ }\textbf {\bibinfo
  {volume} {97}},\ \bibinfo {pages} {245128} (\bibinfo {year}
  {2018})}\BibitemShut {NoStop}%
\bibitem [{\citenamefont {Ghahari}\ \emph {et~al.}(2016)\citenamefont
  {Ghahari}, \citenamefont {Xie}, \citenamefont {Taniguchi}, \citenamefont
  {Watanabe}, \citenamefont {Foster},\ and\ \citenamefont {Kim}}]{ghahari16}%
  \BibitemOpen
  \bibfield  {author} {\bibinfo {author} {\bibfnamefont {F.}~\bibnamefont
  {Ghahari}}, \bibinfo {author} {\bibfnamefont {H.-Y.}\ \bibnamefont {Xie}},
  \bibinfo {author} {\bibfnamefont {T.}~\bibnamefont {Taniguchi}}, \bibinfo
  {author} {\bibfnamefont {K.}~\bibnamefont {Watanabe}}, \bibinfo {author}
  {\bibfnamefont {M.~S.}\ \bibnamefont {Foster}},\ and\ \bibinfo {author}
  {\bibfnamefont {P.}~\bibnamefont {Kim}},\ }\bibfield  {title} {\bibinfo
  {title} {{Enhanced Thermoelectric Power in Graphene: Violation of the Mott
  Relation by Inelastic Scattering}},\ }\href
  {https://doi.org/10.1103/PhysRevLett.116.136802} {\bibfield  {journal}
  {\bibinfo  {journal} {Phys. Rev. Lett.}\ }\textbf {\bibinfo {volume} {116}},\
  \bibinfo {pages} {136802} (\bibinfo {year} {2016})}\BibitemShut {NoStop}%
\bibitem [{\citenamefont {Xie}\ and\ \citenamefont {Foster}(2016)}]{xie16}%
  \BibitemOpen
  \bibfield  {author} {\bibinfo {author} {\bibfnamefont {H.-Y.}\ \bibnamefont
  {Xie}}\ and\ \bibinfo {author} {\bibfnamefont {M.~S.}\ \bibnamefont
  {Foster}},\ }\bibfield  {title} {\bibinfo {title} {{Transport coefficients of
  graphene: Interplay of impurity scattering, Coulomb interaction, and optical
  phonons}},\ }\href {https://doi.org/10.1103/PhysRevB.93.195103} {\bibfield
  {journal} {\bibinfo  {journal} {Phys. Rev. B}\ }\textbf {\bibinfo {volume}
  {93}},\ \bibinfo {pages} {195103} (\bibinfo {year} {2016})}\BibitemShut
  {NoStop}%
\bibitem [{\citenamefont {Boeve}\ and\ \citenamefont
  {Hofmann}(2026)}]{boeve26}%
  \BibitemOpen
  \bibfield  {author} {\bibinfo {author} {\bibfnamefont {S.}~\bibnamefont
  {Boeve}}\ and\ \bibinfo {author} {\bibfnamefont {J.}~\bibnamefont
  {Hofmann}},\ }\href@noop {} {\bibinfo {title} {{Structure and relaxation of
  long-lived odd-parity modes in two-dimensional Fermi liquids}}} (\bibinfo
  {year} {2026}),\ \Eprint {https://arxiv.org/abs/companion paper} {companion
  paper} \BibitemShut {NoStop}%
\bibitem [{\citenamefont {Laikhtman}(1992)}]{laikhtman92}%
  \BibitemOpen
  \bibfield  {author} {\bibinfo {author} {\bibfnamefont {B.}~\bibnamefont
  {Laikhtman}},\ }\bibfield  {title} {\bibinfo {title} {{Electron-electron
  angular relaxation in a two-dimensional electron gas}},\ }\href
  {https://doi.org/10.1103/PhysRevB.45.1259} {\bibfield  {journal} {\bibinfo
  {journal} {Phys. Rev. B}\ }\textbf {\bibinfo {volume} {45}},\ \bibinfo
  {pages} {1259} (\bibinfo {year} {1992})}\BibitemShut {NoStop}%
\bibitem [{\citenamefont {Hodges}\ \emph {et~al.}(1971)\citenamefont {Hodges},
  \citenamefont {Smith},\ and\ \citenamefont {Wilkins}}]{hodges71}%
  \BibitemOpen
  \bibfield  {author} {\bibinfo {author} {\bibfnamefont {C.}~\bibnamefont
  {Hodges}}, \bibinfo {author} {\bibfnamefont {H.}~\bibnamefont {Smith}},\ and\
  \bibinfo {author} {\bibfnamefont {J.~W.}\ \bibnamefont {Wilkins}},\
  }\bibfield  {title} {\bibinfo {title} {{Effect of Fermi Surface Geometry on
  Electron-Electron Scattering}},\ }\href
  {https://doi.org/10.1103/PhysRevB.4.302} {\bibfield  {journal} {\bibinfo
  {journal} {Phys. Rev. B}\ }\textbf {\bibinfo {volume} {4}},\ \bibinfo {pages}
  {302} (\bibinfo {year} {1971})}\BibitemShut {NoStop}%
\bibitem [{\citenamefont {Bloom}(1975)}]{bloom75}%
  \BibitemOpen
  \bibfield  {author} {\bibinfo {author} {\bibfnamefont {P.}~\bibnamefont
  {Bloom}},\ }\bibfield  {title} {\bibinfo {title} {{Two-dimensional Fermi
  gas}},\ }\href {https://doi.org/10.1103/PhysRevB.12.125} {\bibfield
  {journal} {\bibinfo  {journal} {Phys. Rev. B}\ }\textbf {\bibinfo {volume}
  {12}},\ \bibinfo {pages} {125} (\bibinfo {year} {1975})}\BibitemShut
  {NoStop}%
\bibitem [{\citenamefont {Giuliani}\ and\ \citenamefont
  {Quinn}(1982)}]{giuliani82}%
  \BibitemOpen
  \bibfield  {author} {\bibinfo {author} {\bibfnamefont {G.~F.}\ \bibnamefont
  {Giuliani}}\ and\ \bibinfo {author} {\bibfnamefont {J.~J.}\ \bibnamefont
  {Quinn}},\ }\bibfield  {title} {\bibinfo {title} {{Lifetime of a
  quasiparticle in a two-dimensional electron gas}},\ }\href
  {https://doi.org/10.1103/PhysRevB.26.4421} {\bibfield  {journal} {\bibinfo
  {journal} {Phys. Rev. B}\ }\textbf {\bibinfo {volume} {26}},\ \bibinfo
  {pages} {4421} (\bibinfo {year} {1982})}\BibitemShut {NoStop}%
\bibitem [{\citenamefont {Zheng}\ and\ \citenamefont
  {Das~Sarma}(1996)}]{zheng96}%
  \BibitemOpen
  \bibfield  {author} {\bibinfo {author} {\bibfnamefont {L.}~\bibnamefont
  {Zheng}}\ and\ \bibinfo {author} {\bibfnamefont {S.}~\bibnamefont
  {Das~Sarma}},\ }\bibfield  {title} {\bibinfo {title} {{Coulomb scattering
  lifetime of a two-dimensional electron gas}},\ }\href
  {https://doi.org/10.1103/PhysRevB.53.9964} {\bibfield  {journal} {\bibinfo
  {journal} {Phys. Rev. B}\ }\textbf {\bibinfo {volume} {53}},\ \bibinfo
  {pages} {9964} (\bibinfo {year} {1996})}\BibitemShut {NoStop}%
\bibitem [{\citenamefont {Ledwith}\ \emph
  {et~al.}(2019{\natexlab{a}})\citenamefont {Ledwith}, \citenamefont {Guo},\
  and\ \citenamefont {Levitov}}]{ledwith19}%
  \BibitemOpen
  \bibfield  {author} {\bibinfo {author} {\bibfnamefont {P.~J.}\ \bibnamefont
  {Ledwith}}, \bibinfo {author} {\bibfnamefont {H.}~\bibnamefont {Guo}},\ and\
  \bibinfo {author} {\bibfnamefont {L.}~\bibnamefont {Levitov}},\ }\bibfield
  {title} {\bibinfo {title} {{The hierarchy of excitation lifetimes in
  two-dimensional {Fermi} gases}},\ }\href
  {https://doi.org/10.1016/j.aop.2019.167913} {\bibfield  {journal} {\bibinfo
  {journal} {Ann. Phys.}\ }\textbf {\bibinfo {volume} {411}},\ \bibinfo {pages}
  {167913} (\bibinfo {year} {2019}{\natexlab{a}})}\BibitemShut {NoStop}%
\bibitem [{\citenamefont {Hofmann}\ and\ \citenamefont
  {Gran}(2023)}]{hofmann23}%
  \BibitemOpen
  \bibfield  {author} {\bibinfo {author} {\bibfnamefont {J.}~\bibnamefont
  {Hofmann}}\ and\ \bibinfo {author} {\bibfnamefont {U.}~\bibnamefont {Gran}},\
  }\bibfield  {title} {\bibinfo {title} {{Anomalously long lifetimes in
  two-dimensional Fermi liquids}},\ }\href
  {https://doi.org/10.1103/PhysRevB.108.L121401} {\bibfield  {journal}
  {\bibinfo  {journal} {Phys. Rev. B}\ }\textbf {\bibinfo {volume} {108}},\
  \bibinfo {pages} {L121401} (\bibinfo {year} {2023})}\BibitemShut {NoStop}%
\bibitem [{\citenamefont {Nilsson}\ \emph {et~al.}(2025)\citenamefont
  {Nilsson}, \citenamefont {Gran},\ and\ \citenamefont {Hofmann}}]{nilsson25}%
  \BibitemOpen
  \bibfield  {author} {\bibinfo {author} {\bibfnamefont {E.}~\bibnamefont
  {Nilsson}}, \bibinfo {author} {\bibfnamefont {U.}~\bibnamefont {Gran}},\ and\
  \bibinfo {author} {\bibfnamefont {J.}~\bibnamefont {Hofmann}},\ }\bibfield
  {title} {\bibinfo {title} {{Nonequilibrium Relaxation and Odd-Even Effect in
  Finite-Temperature Electron Gases}},\ }\href
  {https://doi.org/10.1103/ly37-5gdw} {\bibfield  {journal} {\bibinfo
  {journal} {Phys. Rev. X}\ }\textbf {\bibinfo {volume} {15}},\ \bibinfo
  {pages} {041007} (\bibinfo {year} {2025})}\BibitemShut {NoStop}%
\bibitem [{\citenamefont {Gurzhi}\ \emph {et~al.}(1982)\citenamefont {Gurzhi},
  \citenamefont {Kopeliovich},\ and\ \citenamefont {Rutkevich}}]{gurzhi82}%
  \BibitemOpen
  \bibfield  {author} {\bibinfo {author} {\bibfnamefont {R.~N.}\ \bibnamefont
  {Gurzhi}}, \bibinfo {author} {\bibfnamefont {A.~I.}\ \bibnamefont
  {Kopeliovich}},\ and\ \bibinfo {author} {\bibfnamefont {S.~B.}\ \bibnamefont
  {Rutkevich}},\ }\bibfield  {title} {\bibinfo {title} {{Electric conductivity
  of two-dimensional metallic systems}},\ }\href
  {https://www.jetp.ras.ru/cgi-bin/e/index/e/56/1/p159?a=list} {\bibfield
  {journal} {\bibinfo  {journal} {Zh. Eksp. Teor. Fiz.}\ }\textbf {\bibinfo
  {volume} {83}},\ \bibinfo {pages} {290} (\bibinfo {year} {1982})},\ \bibinfo
  {note} {[Sov. Phys. JETP \textbf{56}, 159 (1982)]}\BibitemShut {NoStop}%
\bibitem [{\citenamefont {Gurzhi}\ \emph {et~al.}(1987)\citenamefont {Gurzhi},
  \citenamefont {Kopeliovich},\ and\ \citenamefont {Rutkevich}}]{gurzhi87}%
  \BibitemOpen
  \bibfield  {author} {\bibinfo {author} {\bibfnamefont {R.}~\bibnamefont
  {Gurzhi}}, \bibinfo {author} {\bibfnamefont {A.}~\bibnamefont
  {Kopeliovich}},\ and\ \bibinfo {author} {\bibfnamefont {S.}~\bibnamefont
  {Rutkevich}},\ }\bibfield  {title} {\bibinfo {title} {{Kinetic properties of
  two-dimensional metal systems}},\ }\href
  {https://doi.org/10.1080/00018738700101002} {\bibfield  {journal} {\bibinfo
  {journal} {Advances in Physics}\ }\textbf {\bibinfo {volume} {36}},\ \bibinfo
  {pages} {221} (\bibinfo {year} {1987})}\BibitemShut {NoStop}%
\bibitem [{\citenamefont {Gurzhi}\ \emph {et~al.}(1993)\citenamefont {Gurzhi},
  \citenamefont {Kalinenko},\ and\ \citenamefont {Kopeliovich}}]{gurzhi93}%
  \BibitemOpen
  \bibfield  {author} {\bibinfo {author} {\bibfnamefont {R.~N.}\ \bibnamefont
  {Gurzhi}}, \bibinfo {author} {\bibfnamefont {A.~N.}\ \bibnamefont
  {Kalinenko}},\ and\ \bibinfo {author} {\bibfnamefont {A.~I.}\ \bibnamefont
  {Kopeliovich}},\ }\bibfield  {title} {\bibinfo {title} {{On momentum
  relaxation processes during electron collisions in two-dimensional conducting
  systems}},\ }\href {https://doi.org/10.1063/10.0033504} {\bibfield  {journal}
  {\bibinfo  {journal} {Low Temp. Phys.}\ }\textbf {\bibinfo {volume} {19}},\
  \bibinfo {pages} {744} (\bibinfo {year} {1993})},\ \bibinfo {note} {[Fiz.
  Nizk. Temp. \textbf{19}, 1046 (1993)]}\BibitemShut {NoStop}%
\bibitem [{\citenamefont {Gurzhi}\ \emph {et~al.}(1995)\citenamefont {Gurzhi},
  \citenamefont {Kalinenko},\ and\ \citenamefont {Kopeliovich}}]{gurzhi95c}%
  \BibitemOpen
  \bibfield  {author} {\bibinfo {author} {\bibfnamefont {R.~N.}\ \bibnamefont
  {Gurzhi}}, \bibinfo {author} {\bibfnamefont {A.~N.}\ \bibnamefont
  {Kalinenko}},\ and\ \bibinfo {author} {\bibfnamefont {A.~I.}\ \bibnamefont
  {Kopeliovich}},\ }\bibfield  {title} {\bibinfo {title} {Electron-electron
  collisions and a new hydrodynamic effect in two-dimensional electron gas},\
  }\href {https://doi.org/10.1103/PhysRevLett.74.3872} {\bibfield  {journal}
  {\bibinfo  {journal} {Phys. Rev. Lett.}\ }\textbf {\bibinfo {volume} {74}},\
  \bibinfo {pages} {3872} (\bibinfo {year} {1995})}\BibitemShut {NoStop}%
\bibitem [{\citenamefont {Ledwith}\ \emph
  {et~al.}(2019{\natexlab{b}})\citenamefont {Ledwith}, \citenamefont {Guo},
  \citenamefont {Shytov},\ and\ \citenamefont {Levitov}}]{ledwith19b}%
  \BibitemOpen
  \bibfield  {author} {\bibinfo {author} {\bibfnamefont {P.}~\bibnamefont
  {Ledwith}}, \bibinfo {author} {\bibfnamefont {H.}~\bibnamefont {Guo}},
  \bibinfo {author} {\bibfnamefont {A.}~\bibnamefont {Shytov}},\ and\ \bibinfo
  {author} {\bibfnamefont {L.}~\bibnamefont {Levitov}},\ }\bibfield  {title}
  {\bibinfo {title} {{Tomographic Dynamics and Scale-Dependent Viscosity in 2D
  Electron Systems}},\ }\href {https://doi.org/10.1103/PhysRevLett.123.116601}
  {\bibfield  {journal} {\bibinfo  {journal} {Phys. Rev. Lett.}\ }\textbf
  {\bibinfo {volume} {123}},\ \bibinfo {pages} {116601} (\bibinfo {year}
  {2019}{\natexlab{b}})}\BibitemShut {NoStop}%
\bibitem [{\citenamefont {Hofmann}\ and\ \citenamefont
  {Das~Sarma}(2022)}]{hofmann22a}%
  \BibitemOpen
  \bibfield  {author} {\bibinfo {author} {\bibfnamefont {J.}~\bibnamefont
  {Hofmann}}\ and\ \bibinfo {author} {\bibfnamefont {S.}~\bibnamefont
  {Das~Sarma}},\ }\bibfield  {title} {\bibinfo {title} {{Collective modes in
  interacting two-dimensional tomographic Fermi liquids}},\ }\href
  {https://doi.org/10.1103/PhysRevB.106.205412} {\bibfield  {journal} {\bibinfo
   {journal} {Phys. Rev. B}\ }\textbf {\bibinfo {volume} {106}},\ \bibinfo
  {pages} {205412} (\bibinfo {year} {2022})}\BibitemShut {NoStop}%
\bibitem [{\citenamefont {Kryhin}\ and\ \citenamefont
  {Levitov}(2023)}]{kryhin23}%
  \BibitemOpen
  \bibfield  {author} {\bibinfo {author} {\bibfnamefont {S.}~\bibnamefont
  {Kryhin}}\ and\ \bibinfo {author} {\bibfnamefont {L.}~\bibnamefont
  {Levitov}},\ }\bibfield  {title} {\bibinfo {title} {{Collinear scattering and
  long-lived excitations in two-dimensional electron fluids}},\ }\href
  {https://doi.org/10.1103/PhysRevB.107.L201404} {\bibfield  {journal}
  {\bibinfo  {journal} {Phys. Rev. B}\ }\textbf {\bibinfo {volume} {107}},\
  \bibinfo {pages} {L201404} (\bibinfo {year} {2023})}\BibitemShut {NoStop}%
\bibitem [{\citenamefont {Kryhin}\ \emph {et~al.}(2025)\citenamefont {Kryhin},
  \citenamefont {Hong},\ and\ \citenamefont {Levitov}}]{kryhin25}%
  \BibitemOpen
  \bibfield  {author} {\bibinfo {author} {\bibfnamefont {S.}~\bibnamefont
  {Kryhin}}, \bibinfo {author} {\bibfnamefont {Q.}~\bibnamefont {Hong}},\ and\
  \bibinfo {author} {\bibfnamefont {L.}~\bibnamefont {Levitov}},\ }\bibfield
  {title} {\bibinfo {title} {{Linear-in-temperature conductance in
  two-dimensional electron fluids}},\ }\href
  {https://doi.org/10.1103/PhysRevB.111.L081403} {\bibfield  {journal}
  {\bibinfo  {journal} {Phys. Rev. B}\ }\textbf {\bibinfo {volume} {111}},\
  \bibinfo {pages} {L081403} (\bibinfo {year} {2025})}\BibitemShut {NoStop}%
\bibitem [{\citenamefont {Estrada-{\'A}lvarez}\ \emph
  {et~al.}(2025)\citenamefont {Estrada-{\'A}lvarez}, \citenamefont
  {D{\'{\i}}az},\ and\ \citenamefont {Dom{\'{\i}}nguez-Adame}}]{alvarez25}%
  \BibitemOpen
  \bibfield  {author} {\bibinfo {author} {\bibfnamefont {J.}~\bibnamefont
  {Estrada-{\'A}lvarez}}, \bibinfo {author} {\bibfnamefont {E.}~\bibnamefont
  {D{\'{\i}}az}},\ and\ \bibinfo {author} {\bibfnamefont {F.}~\bibnamefont
  {Dom{\'{\i}}nguez-Adame}},\ }\bibfield  {title} {\bibinfo {title}
  {{Superballistic Paradox in Electron Fluids: Relevance of Tomographic
  Transport}},\ }\href {https://doi.org/10.1103/xsjh-lq5h} {\bibfield
  {journal} {\bibinfo  {journal} {Phys. Rev. Lett.}\ }\textbf {\bibinfo
  {volume} {135}},\ \bibinfo {pages} {206301} (\bibinfo {year}
  {2025})}\BibitemShut {NoStop}%
\bibitem [{\citenamefont {Starkov}\ and\ \citenamefont
  {Trauzettel}(2026)}]{starkov26a}%
  \BibitemOpen
  \bibfield  {author} {\bibinfo {author} {\bibfnamefont {G.~A.}\ \bibnamefont
  {Starkov}}\ and\ \bibinfo {author} {\bibfnamefont {B.}~\bibnamefont
  {Trauzettel}},\ }\bibfield  {title} {\bibinfo {title} {{Anomalous Knudsen
  effect signaling long-lived modes in two-dimensional electron gases}},\
  }\href {https://doi.org/10.1103/rwjf-lh2r} {\bibfield  {journal} {\bibinfo
  {journal} {Phys. Rev. B}\ }\textbf {\bibinfo {volume} {113}},\ \bibinfo
  {pages} {L041406} (\bibinfo {year} {2026})}\BibitemShut {NoStop}%
\bibitem [{\citenamefont {Ben-Shachar}\ and\ \citenamefont
  {Hofmann}(2026{\natexlab{a}})}]{benshachar25c}%
  \BibitemOpen
  \bibfield  {author} {\bibinfo {author} {\bibfnamefont {N.}~\bibnamefont
  {Ben-Shachar}}\ and\ \bibinfo {author} {\bibfnamefont {J.}~\bibnamefont
  {Hofmann}},\ }\bibfield  {title} {\bibinfo {title} {{Tomographic Electron
  Flow in Confined Geometries: Beyond the Dual-Relaxation-Time
  Approximation}},\ }\href {https://doi.org/10.1103/m7cy-fxy9} {\bibfield
  {journal} {\bibinfo  {journal} {Phys. Rev. X}\ }\textbf {\bibinfo {volume}
  {16}},\ \bibinfo {pages} {041044} (\bibinfo {year}
  {2026}{\natexlab{a}})}\BibitemShut {NoStop}%
\bibitem [{\citenamefont {Farrell}\ and\ \citenamefont
  {Lucas}(2026)}]{farrell26}%
  \BibitemOpen
  \bibfield  {author} {\bibinfo {author} {\bibfnamefont {J.~H.}\ \bibnamefont
  {Farrell}}\ and\ \bibinfo {author} {\bibfnamefont {A.}~\bibnamefont
  {Lucas}},\ }\bibfield  {title} {\bibinfo {title} {{Characterizing electronic
  scattering rates with transport in multiterminal devices}},\ }\href
  {https://doi.org/10.1103/69kc-gdwv} {\bibfield  {journal} {\bibinfo
  {journal} {Phys. Rev. B}\ }\textbf {\bibinfo {volume} {114}},\ \bibinfo
  {pages} {295113} (\bibinfo {year} {2026})}\BibitemShut {NoStop}%
\bibitem [{\citenamefont {Ben-Shachar}\ and\ \citenamefont
  {Hofmann}(2026{\natexlab{b}})}]{benshachar26}%
  \BibitemOpen
  \bibfield  {author} {\bibinfo {author} {\bibfnamefont {N.}~\bibnamefont
  {Ben-Shachar}}\ and\ \bibinfo {author} {\bibfnamefont {J.}~\bibnamefont
  {Hofmann}},\ }\bibfield  {title} {\bibinfo {title} {{Magnetotransport of
  tomographic electrons in a channel}},\ }\bibfield  {journal} {\bibinfo
  {journal} {Communications Physics}\ }\href
  {https://doi.org/10.1038/s42005-026-02786-2} {10.1038/s42005-026-02786-2}
  (\bibinfo {year} {2026}{\natexlab{b}})\BibitemShut {NoStop}%
\bibitem [{\citenamefont {Rostami}\ \emph {et~al.}(2025)\citenamefont
  {Rostami}, \citenamefont {Ben-Shachar}, \citenamefont {Moroz},\ and\
  \citenamefont {Hofmann}}]{rostami25}%
  \BibitemOpen
  \bibfield  {author} {\bibinfo {author} {\bibfnamefont {H.}~\bibnamefont
  {Rostami}}, \bibinfo {author} {\bibfnamefont {N.}~\bibnamefont
  {Ben-Shachar}}, \bibinfo {author} {\bibfnamefont {S.}~\bibnamefont {Moroz}},\
  and\ \bibinfo {author} {\bibfnamefont {J.}~\bibnamefont {Hofmann}},\
  }\bibfield  {title} {\bibinfo {title} {{Magnetic field suppression of
  tomographic electron transport}},\ }\href
  {https://doi.org/10.1103/PhysRevB.111.155434} {\bibfield  {journal} {\bibinfo
   {journal} {Phys. Rev. B}\ }\textbf {\bibinfo {volume} {111}},\ \bibinfo
  {pages} {155434} (\bibinfo {year} {2025})}\BibitemShut {NoStop}%
\bibitem [{\citenamefont {Maki}\ and\ \citenamefont {Hofmann}(2026)}]{maki26}%
  \BibitemOpen
  \bibfield  {author} {\bibinfo {author} {\bibfnamefont {J.}~\bibnamefont
  {Maki}}\ and\ \bibinfo {author} {\bibfnamefont {J.}~\bibnamefont {Hofmann}},\
  }\bibfield  {title} {\bibinfo {title} {{Tomographic collective modes in a
  magnetic field}},\ }\href {https://doi.org/10.1103/kfqw-qlsg} {\bibfield
  {journal} {\bibinfo  {journal} {Phys. Rev. B}\ }\textbf {\bibinfo {volume}
  {113}},\ \bibinfo {pages} {245122} (\bibinfo {year} {2026})}\BibitemShut
  {NoStop}%
\bibitem [{\citenamefont {Starkov}(2026)}]{starkov26b}%
  \BibitemOpen
  \bibfield  {author} {\bibinfo {author} {\bibfnamefont {G.~A.}\ \bibnamefont
  {Starkov}},\ }\bibfield  {title} {\bibinfo {title} {{Tomographic flow regime
  vs even-odd effect for the magnetotransport in the Corbino geometry}},\
  }\bibfield  {journal} {\bibinfo  {journal} {arXiv:2607.14929}\ }\href
  {https://doi.org/10.48550/arXiv.2607.14929} {10.48550/arXiv.2607.14929}
  (\bibinfo {year} {2026})\BibitemShut {NoStop}%
\bibitem [{\citenamefont {Ben-Shachar}\ and\ \citenamefont
  {Hofmann}(2026{\natexlab{c}})}]{benshachar26b}%
  \BibitemOpen
  \bibfield  {author} {\bibinfo {author} {\bibfnamefont {N.}~\bibnamefont
  {Ben-Shachar}}\ and\ \bibinfo {author} {\bibfnamefont {J.}~\bibnamefont
  {Hofmann}},\ }\href {https://doi.org/10.48550/arXiv.2607.05540} {\bibinfo
  {title} {{Magnetotransport of tomographic electrons in a Corbino disk}}}
  (\bibinfo {year} {2026}{\natexlab{c}}),\ \Eprint
  {https://arxiv.org/abs/2607.05540} {arXiv:2607.05540} \BibitemShut {NoStop}%
\bibitem [{\citenamefont {Moiseenko}\ \emph {et~al.}(2025)\citenamefont
  {Moiseenko}, \citenamefont {M{\"o}nch}, \citenamefont {Kapralov},
  \citenamefont {Bandurin}, \citenamefont {Ganichev},\ and\ \citenamefont
  {Svintsov}}]{moiseenko25}%
  \BibitemOpen
  \bibfield  {author} {\bibinfo {author} {\bibfnamefont {I.}~\bibnamefont
  {Moiseenko}}, \bibinfo {author} {\bibfnamefont {E.}~\bibnamefont
  {M{\"o}nch}}, \bibinfo {author} {\bibfnamefont {K.}~\bibnamefont {Kapralov}},
  \bibinfo {author} {\bibfnamefont {D.}~\bibnamefont {Bandurin}}, \bibinfo
  {author} {\bibfnamefont {S.}~\bibnamefont {Ganichev}},\ and\ \bibinfo
  {author} {\bibfnamefont {D.}~\bibnamefont {Svintsov}},\ }\bibfield  {title}
  {\bibinfo {title} {{Testing the Tomographic Fermi Liquid Hypothesis with
  High-Order Cyclotron Resonance}},\ }\href {https://doi.org/10.1103/zq1g-8d4s}
  {\bibfield  {journal} {\bibinfo  {journal} {Phys. Rev. Lett.}\ }\textbf
  {\bibinfo {volume} {134}},\ \bibinfo {pages} {226902} (\bibinfo {year}
  {2025})}\BibitemShut {NoStop}%
\bibitem [{\citenamefont {Thuillier}\ and\ \citenamefont
  {Scaffidi}(2026)}]{thuillier26}%
  \BibitemOpen
  \bibfield  {author} {\bibinfo {author} {\bibfnamefont {D.}~\bibnamefont
  {Thuillier}}\ and\ \bibinfo {author} {\bibfnamefont {T.}~\bibnamefont
  {Scaffidi}},\ }\bibfield  {title} {\bibinfo {title} {{ac Fingerprints of 2D
  Electron Hydrodynamics: Superdiffusion and Drude Weight Suppression}},\
  }\href {https://doi.org/10.1103/p8sw-99rn} {\bibfield  {journal} {\bibinfo
  {journal} {Phys. Rev. Lett.}\ }\textbf {\bibinfo {volume} {137}},\ \bibinfo
  {pages} {196303} (\bibinfo {year} {2026})}\BibitemShut {NoStop}%
\bibitem [{\citenamefont {Callaway}(1959)}]{callaway59}%
  \BibitemOpen
  \bibfield  {author} {\bibinfo {author} {\bibfnamefont {J.}~\bibnamefont
  {Callaway}},\ }\bibfield  {title} {\bibinfo {title} {{Model for Lattice
  Thermal Conductivity at Low Temperatures}},\ }\href
  {https://doi.org/10.1103/PhysRev.113.1046} {\bibfield  {journal} {\bibinfo
  {journal} {Phys. Rev.}\ }\textbf {\bibinfo {volume} {113}},\ \bibinfo {pages}
  {1046} (\bibinfo {year} {1959})}\BibitemShut {NoStop}%
\bibitem [{\citenamefont {de~Jong}\ and\ \citenamefont
  {Molenkamp}(1995)}]{dejong95}%
  \BibitemOpen
  \bibfield  {author} {\bibinfo {author} {\bibfnamefont {M.~J.~M.}\
  \bibnamefont {de~Jong}}\ and\ \bibinfo {author} {\bibfnamefont {L.~W.}\
  \bibnamefont {Molenkamp}},\ }\bibfield  {title} {\bibinfo {title}
  {{Hydrodynamic Electron Flow in High-Mobility Wires}},\ }\href
  {https://doi.org/10.1103/PhysRevB.51.13389} {\bibfield  {journal} {\bibinfo
  {journal} {Phys. Rev. B}\ }\textbf {\bibinfo {volume} {51}},\ \bibinfo
  {pages} {13389} (\bibinfo {year} {1995})}\BibitemShut {NoStop}%
\bibitem [{\citenamefont {Zeng}\ \emph {et~al.}(2026)\citenamefont {Zeng},
  \citenamefont {Guo}, \citenamefont {Ghosh}, \citenamefont {Watanabe},
  \citenamefont {Taniguchi}, \citenamefont {Levitov},\ and\ \citenamefont
  {Dean}}]{zeng24}%
  \BibitemOpen
  \bibfield  {author} {\bibinfo {author} {\bibfnamefont {Y.}~\bibnamefont
  {Zeng}}, \bibinfo {author} {\bibfnamefont {H.}~\bibnamefont {Guo}}, \bibinfo
  {author} {\bibfnamefont {O.~M.}\ \bibnamefont {Ghosh}}, \bibinfo {author}
  {\bibfnamefont {K.}~\bibnamefont {Watanabe}}, \bibinfo {author}
  {\bibfnamefont {T.}~\bibnamefont {Taniguchi}}, \bibinfo {author}
  {\bibfnamefont {L.~S.}\ \bibnamefont {Levitov}},\ and\ \bibinfo {author}
  {\bibfnamefont {C.~R.}\ \bibnamefont {Dean}},\ }\bibfield  {title} {\bibinfo
  {title} {{Quantitative measurement of viscosity in two-dimensional electron
  fluids}},\ }\href {https://doi.org/10.1126/science.adr5469} {\bibfield
  {journal} {\bibinfo  {journal} {Science}\ }\textbf {\bibinfo {volume}
  {394}},\ \bibinfo {pages} {786} (\bibinfo {year} {2026})}\BibitemShut
  {NoStop}%
\bibitem [{\citenamefont {Risken}(1996)}]{risken96}%
  \BibitemOpen
  \bibfield  {author} {\bibinfo {author} {\bibfnamefont {H.}~\bibnamefont
  {Risken}},\ }\href {https://doi.org/10.1007/978-3-642-61544-3} {\emph
  {\bibinfo {title} {{The Fokker-Planck Equation: Methods of Solution and
  Applications}}}},\ \bibinfo {edition} {2nd}\ ed.\ (\bibinfo  {publisher}
  {Springer},\ \bibinfo {address} {Berlin},\ \bibinfo {year} {1996})\ \bibinfo
  {note} {{Chap.~9}}\BibitemShut {NoStop}%
\bibitem [{\citenamefont {Nazaryan}\ and\ \citenamefont
  {Levitov}(2024)}]{nazaryan24}%
  \BibitemOpen
  \bibfield  {author} {\bibinfo {author} {\bibfnamefont {K.~G.}\ \bibnamefont
  {Nazaryan}}\ and\ \bibinfo {author} {\bibfnamefont {L.}~\bibnamefont
  {Levitov}},\ }\bibfield  {title} {\bibinfo {title} {{Nonlocal conductivity,
  continued fractions, and current vortices in electron fluids}},\ }\href
  {https://doi.org/10.1103/PhysRevB.110.045147} {\bibfield  {journal} {\bibinfo
   {journal} {Phys. Rev. B}\ }\textbf {\bibinfo {volume} {110}},\ \bibinfo
  {pages} {045147} (\bibinfo {year} {2024})}\BibitemShut {NoStop}%
\bibitem [{\citenamefont {Behnia}\ \emph {et~al.}(2004)\citenamefont {Behnia},
  \citenamefont {Jaccard},\ and\ \citenamefont {Flouquet}}]{behnia04}%
  \BibitemOpen
  \bibfield  {author} {\bibinfo {author} {\bibfnamefont {K.}~\bibnamefont
  {Behnia}}, \bibinfo {author} {\bibfnamefont {D.}~\bibnamefont {Jaccard}},\
  and\ \bibinfo {author} {\bibfnamefont {J.}~\bibnamefont {Flouquet}},\
  }\bibfield  {title} {\bibinfo {title} {{On the thermoelectricity of
  correlated electrons in the zero-temperature limit}},\ }\href
  {https://doi.org/10.1088/0953-8984/16/28/037} {\bibfield  {journal} {\bibinfo
   {journal} {Journal of Physics: Condensed Matter}\ }\textbf {\bibinfo
  {volume} {16}},\ \bibinfo {pages} {5187} (\bibinfo {year}
  {2004})}\BibitemShut {NoStop}%
\bibitem [{\citenamefont {Chickering}\ \emph {et~al.}(2009)\citenamefont
  {Chickering}, \citenamefont {Eisenstein},\ and\ \citenamefont
  {Reno}}]{chickering09}%
  \BibitemOpen
  \bibfield  {author} {\bibinfo {author} {\bibfnamefont {W.~E.}\ \bibnamefont
  {Chickering}}, \bibinfo {author} {\bibfnamefont {J.~P.}\ \bibnamefont
  {Eisenstein}},\ and\ \bibinfo {author} {\bibfnamefont {J.~L.}\ \bibnamefont
  {Reno}},\ }\bibfield  {title} {\bibinfo {title} {{Hot-Electron Thermocouple
  and the Diffusion Thermopower of Two-Dimensional Electrons in GaAs}},\ }\href
  {https://doi.org/10.1103/PhysRevLett.103.046807} {\bibfield  {journal}
  {\bibinfo  {journal} {Phys. Rev. Lett.}\ }\textbf {\bibinfo {volume} {103}},\
  \bibinfo {pages} {046807} (\bibinfo {year} {2009})}\BibitemShut {NoStop}%
\bibitem [{\citenamefont {Pusep}\ \emph {et~al.}(2026)\citenamefont {Pusep},
  \citenamefont {Patricio}, \citenamefont {Glazov}, \citenamefont {Oliveira},
  \citenamefont {Teodoro}, \citenamefont {Levin}, \citenamefont {Bakarov},\
  and\ \citenamefont {Gusev}}]{pusep26}%
  \BibitemOpen
  \bibfield  {author} {\bibinfo {author} {\bibfnamefont {Y.~A.}\ \bibnamefont
  {Pusep}}, \bibinfo {author} {\bibfnamefont {M.~A.~T.}\ \bibnamefont
  {Patricio}}, \bibinfo {author} {\bibfnamefont {M.~M.}\ \bibnamefont
  {Glazov}}, \bibinfo {author} {\bibfnamefont {V.~A.}\ \bibnamefont
  {Oliveira}}, \bibinfo {author} {\bibfnamefont {M.~D.}\ \bibnamefont
  {Teodoro}}, \bibinfo {author} {\bibfnamefont {A.~D.}\ \bibnamefont {Levin}},
  \bibinfo {author} {\bibfnamefont {A.~K.}\ \bibnamefont {Bakarov}},\ and\
  \bibinfo {author} {\bibfnamefont {G.~M.}\ \bibnamefont {Gusev}},\ }\bibfield
  {title} {\bibinfo {title} {{Spectroscopy of heat transport and violation of
  the Wiedemann-Franz law in GaAs hydrodynamic mesoscopic channel}},\ }\href
  {https://doi.org/10.1038/s41598-026-45858-7} {\bibfield  {journal} {\bibinfo
  {journal} {Scientific Reports}\ }\textbf {\bibinfo {volume} {16}},\ \bibinfo
  {pages} {17385} (\bibinfo {year} {2026})}\BibitemShut {NoStop}%
\bibitem [{\citenamefont {Wang}\ \emph {et~al.}(2022)\citenamefont {Wang},
  \citenamefont {Jia}, \citenamefont {Du}, \citenamefont {Pfeiffer},
  \citenamefont {Baldwin},\ and\ \citenamefont {West}}]{wang22}%
  \BibitemOpen
  \bibfield  {author} {\bibinfo {author} {\bibfnamefont {X.}~\bibnamefont
  {Wang}}, \bibinfo {author} {\bibfnamefont {P.}~\bibnamefont {Jia}}, \bibinfo
  {author} {\bibfnamefont {R.-R.}\ \bibnamefont {Du}}, \bibinfo {author}
  {\bibfnamefont {L.~N.}\ \bibnamefont {Pfeiffer}}, \bibinfo {author}
  {\bibfnamefont {K.~W.}\ \bibnamefont {Baldwin}},\ and\ \bibinfo {author}
  {\bibfnamefont {K.~W.}\ \bibnamefont {West}},\ }\bibfield  {title} {\bibinfo
  {title} {{Hydrodynamic charge transport in an GaAs/AlGaAs ultrahigh-mobility
  two-dimensional electron gas}},\ }\href
  {https://doi.org/10.1103/PhysRevB.106.L241302} {\bibfield  {journal}
  {\bibinfo  {journal} {Phys. Rev. B}\ }\textbf {\bibinfo {volume} {106}},\
  \bibinfo {pages} {L241302} (\bibinfo {year} {2022})}\BibitemShut {NoStop}%
\bibitem [{\citenamefont {Chung}\ \emph {et~al.}(2022)\citenamefont {Chung},
  \citenamefont {Gupta}, \citenamefont {Baldwin}, \citenamefont {West},
  \citenamefont {Shayegan},\ and\ \citenamefont {Pfeiffer}}]{chung22}%
  \BibitemOpen
  \bibfield  {author} {\bibinfo {author} {\bibfnamefont {Y.~J.}\ \bibnamefont
  {Chung}}, \bibinfo {author} {\bibfnamefont {A.}~\bibnamefont {Gupta}},
  \bibinfo {author} {\bibfnamefont {K.~W.}\ \bibnamefont {Baldwin}}, \bibinfo
  {author} {\bibfnamefont {K.~W.}\ \bibnamefont {West}}, \bibinfo {author}
  {\bibfnamefont {M.}~\bibnamefont {Shayegan}},\ and\ \bibinfo {author}
  {\bibfnamefont {L.~N.}\ \bibnamefont {Pfeiffer}},\ }\bibfield  {title}
  {\bibinfo {title} {{Understanding limits to mobility in ultrahigh-mobility
  GaAs two-dimensional electron systems: 100 million
  ${\mathrm{cm}}^{2}/\mathrm{Vs}$ and beyond}},\ }\href
  {https://doi.org/10.1103/PhysRevB.106.075134} {\bibfield  {journal} {\bibinfo
   {journal} {Phys. Rev. B}\ }\textbf {\bibinfo {volume} {106}},\ \bibinfo
  {pages} {075134} (\bibinfo {year} {2022})}\BibitemShut {NoStop}%
\bibitem [{\citenamefont {Das~Sarma}\ and\ \citenamefont
  {Stern}(1985)}]{dassarma85}%
  \BibitemOpen
  \bibfield  {author} {\bibinfo {author} {\bibfnamefont {S.}~\bibnamefont
  {Das~Sarma}}\ and\ \bibinfo {author} {\bibfnamefont {F.}~\bibnamefont
  {Stern}},\ }\bibfield  {title} {\bibinfo {title} {{Single-particle relaxation
  time versus scattering time in an impure electron gas}},\ }\href
  {https://doi.org/10.1103/PhysRevB.32.8442} {\bibfield  {journal} {\bibinfo
  {journal} {Phys. Rev. B}\ }\textbf {\bibinfo {volume} {32}},\ \bibinfo
  {pages} {8442(R)} (\bibinfo {year} {1985})}\BibitemShut {NoStop}%
\bibitem [{\citenamefont {MacLeod}\ \emph {et~al.}(2009)\citenamefont
  {MacLeod}, \citenamefont {Chan}, \citenamefont {Martin}, \citenamefont
  {Hamilton}, \citenamefont {See}, \citenamefont {Micolich}, \citenamefont
  {Aagesen},\ and\ \citenamefont {Lindelof}}]{macleod09}%
  \BibitemOpen
  \bibfield  {author} {\bibinfo {author} {\bibfnamefont {S.~J.}\ \bibnamefont
  {MacLeod}}, \bibinfo {author} {\bibfnamefont {K.}~\bibnamefont {Chan}},
  \bibinfo {author} {\bibfnamefont {T.~P.}\ \bibnamefont {Martin}}, \bibinfo
  {author} {\bibfnamefont {A.~R.}\ \bibnamefont {Hamilton}}, \bibinfo {author}
  {\bibfnamefont {A.}~\bibnamefont {See}}, \bibinfo {author} {\bibfnamefont
  {A.~P.}\ \bibnamefont {Micolich}}, \bibinfo {author} {\bibfnamefont
  {M.}~\bibnamefont {Aagesen}},\ and\ \bibinfo {author} {\bibfnamefont {P.~E.}\
  \bibnamefont {Lindelof}},\ }\bibfield  {title} {\bibinfo {title} {{Role of
  background impurities in the single-particle relaxation lifetime of a
  two-dimensional electron gas}},\ }\href
  {https://doi.org/10.1103/PhysRevB.80.035310} {\bibfield  {journal} {\bibinfo
  {journal} {Phys. Rev. B}\ }\textbf {\bibinfo {volume} {80}},\ \bibinfo
  {pages} {035310} (\bibinfo {year} {2009})}\BibitemShut {NoStop}%
\bibitem [{\citenamefont {Brantut}\ \emph {et~al.}(2013)\citenamefont
  {Brantut}, \citenamefont {Grenier}, \citenamefont {Meineke}, \citenamefont
  {Stadler}, \citenamefont {Krinner}, \citenamefont {Kollath}, \citenamefont
  {Esslinger},\ and\ \citenamefont {Georges}}]{brantut13}%
  \BibitemOpen
  \bibfield  {author} {\bibinfo {author} {\bibfnamefont {J.-P.}\ \bibnamefont
  {Brantut}}, \bibinfo {author} {\bibfnamefont {C.}~\bibnamefont {Grenier}},
  \bibinfo {author} {\bibfnamefont {J.}~\bibnamefont {Meineke}}, \bibinfo
  {author} {\bibfnamefont {D.}~\bibnamefont {Stadler}}, \bibinfo {author}
  {\bibfnamefont {S.}~\bibnamefont {Krinner}}, \bibinfo {author} {\bibfnamefont
  {C.}~\bibnamefont {Kollath}}, \bibinfo {author} {\bibfnamefont
  {T.}~\bibnamefont {Esslinger}},\ and\ \bibinfo {author} {\bibfnamefont
  {A.}~\bibnamefont {Georges}},\ }\bibfield  {title} {\bibinfo {title} {{A
  Thermoelectric Heat Engine with Ultracold Atoms}},\ }\href
  {https://doi.org/10.1126/science.1242308} {\bibfield  {journal} {\bibinfo
  {journal} {Science}\ }\textbf {\bibinfo {volume} {342}},\ \bibinfo {pages}
  {713} (\bibinfo {year} {2013})}\BibitemShut {NoStop}%
\bibitem [{\citenamefont {Maki}\ \emph {et~al.}(2025)\citenamefont {Maki},
  \citenamefont {Gran},\ and\ \citenamefont {Hofmann}}]{maki25}%
  \BibitemOpen
  \bibfield  {author} {\bibinfo {author} {\bibfnamefont {J.}~\bibnamefont
  {Maki}}, \bibinfo {author} {\bibfnamefont {U.}~\bibnamefont {Gran}},\ and\
  \bibinfo {author} {\bibfnamefont {J.}~\bibnamefont {Hofmann}},\ }\bibfield
  {title} {\bibinfo {title} {{Odd-parity effect and scale-dependent viscosity
  in atomic quantum gases}},\ }\href
  {https://doi.org/10.1038/s42005-025-02231-w} {\bibfield  {journal} {\bibinfo
  {journal} {Communications Physics}\ }\textbf {\bibinfo {volume} {8}},\
  \bibinfo {pages} {319} (\bibinfo {year} {2025})}\BibitemShut {NoStop}%
\bibitem [{\citenamefont {Beardo}\ \emph {et~al.}(2026)\citenamefont {Beardo},
  \citenamefont {Ramos-Martín}, \citenamefont {Tur-Prats}, \citenamefont
  {Alvarez},\ and\ \citenamefont {Camacho}}]{beardo26}%
  \BibitemOpen
  \bibfield  {author} {\bibinfo {author} {\bibfnamefont {A.}~\bibnamefont
  {Beardo}}, \bibinfo {author} {\bibfnamefont {A.}~\bibnamefont
  {Ramos-Martín}}, \bibinfo {author} {\bibfnamefont {J.}~\bibnamefont
  {Tur-Prats}}, \bibinfo {author} {\bibfnamefont {F.~X.}\ \bibnamefont
  {Alvarez}},\ and\ \bibinfo {author} {\bibfnamefont {J.}~\bibnamefont
  {Camacho}},\ }\href {https://doi.org/10.48550/arXiv.2610.32272} {\bibinfo
  {title} {{Nonlocal Thermoelectric Transport in Degenerate Electron Systems}}}
  (\bibinfo {year} {2026}),\ \Eprint {https://arxiv.org/abs/2610.32272}
  {arXiv:2610.32272} \BibitemShut {NoStop}%
\end{thebibliography}%

\appendix

\section{END MATTER}

\paragraph{Collision integral in the two-moment basis.---} We parametrize the two-moment collision matrix in the form
\begin{align}
    L_m &= \frac{W_c T^2}{\hbar T_F} 
    \begin{pmatrix} 
    g_m + G_m k_m^2 & - G_m k_m \\ 
    - G_m k_m & G_m 
    \end{pmatrix} , \label{eq:collisionmatrix}
\end{align}
where, in terms of the collision-matrix elements, $g_m = L_{00} - L_{01}^2/L_{11}$, $G_m=L_{11}$, and $k_m = - L_{01}/L_{11}$. $W_c$ is the amplitude of the constant interaction. We define $\gamma_\eta$ as the smallest eigenvalue of $L_{m=2}$. For odd $m$, to leading order at low temperature, the collision matrix has eigenvector $(1,k_m)$ and eigenvalue $g_m$, which is the long-lived mode, and eigenvector $(-k_m,1)$ with eigenvalue $G_m$, which is a fast mode that relaxes by collinear scattering. In the parametrization of Eq.~\eqref{eq:collisionmatrix}, the parameters $G_m$ and $g_m$ describe the lifetimes of modes, while $k_m = \kappa \vartheta_m$ sets their tilt. They can be taken from numerical simulations or kept as general model parameters. Here, we choose the explicit low-temperature form for a constant interaction, which is valid for $m^2T/T_F \lesssim 0.4$~\cite{boeve26}:
\begin{align}
    G_m &= 
    \begin{cases}
        \frac{24\pi}{5} o_m & m \ {\rm even} \\
        \frac{8\pi}{5} \bar{\Lambda}_T & m \ {\rm odd}
    \end{cases}
\end{align}
with $o_m$ defined after Eq.~\eqref{eq:xim}. Furthermore,
\begin{align}
    g_m &= 
    \begin{cases}
        \frac{8 \pi}{3} o_m - G_m k_m^2 & m \ {\rm even} \\
        \frac{2\pi^3}{15} m^4 \frac{T^2}{T_F^2} (1 - \frac{1}{m^2} + \bar{\Lambda}_T \xi_m (1-\xi_m)) & m \ {\rm odd} ,
    \end{cases} \label{eq:defgm}
\end{align}
where $\xi_m$ is defined in Eq.~\eqref{eq:xim}. We have $g_{m=1}=0$, as required for the boost. Since the odd-mode rate at large $m$ approaches the Fermi-liquid scaling~\cite{nilsson25}, in our calculations we replace $g_m$ for odd $m$ by $G_m$ if it exceeds $G_m$. Finally, $k_m=\kappa \vartheta_m$ with tilt
\begin{align}
    \vartheta_m &= \begin{cases}
        - 
        \frac{\bar{\Lambda}_T}{3o_m} m^2 \tilde{\xi}_m & m \ {\rm even} \\
        m^2 \xi_m & m \ {\rm odd} ,
    \end{cases} \label{eq:varthetam}
\end{align}
where $\tilde{\xi}_m = 1 - 2 (\tilde{a}_m-1)/\bar{\Lambda}_T$ with $\tilde{a}_m = (2+1/m^2) o_m$ is the relative even-mode dressing. Since $\xi_m$ and $\tilde{\xi}_m$ cross zero and become negative above $m^2 T/T_F \simeq 0.55$, we replace them for $m\geq 2$ in both Eqs.~\eqref{eq:defgm} and~\eqref{eq:varthetam} by the soft zero
\begin{align}
    \xi_m^{\rm int} &= \frac{1}{\zeta \bar{\Lambda}_T} \ln \bigl(1 + e^{\zeta \bar{\Lambda}_T \xi_m}\bigr) 
\end{align}
with fit parameter $\zeta = 2$, and likewise for $\tilde{\xi}_m$, which also ensures that $g_m$ is positive. This interpolation agrees with exact-diagonalization results of the thermal dressing at all temperatures~\cite{boeve26}. We note that we use the constant interaction to provide simple closed-form expressions for our model definition. A screened Coulomb interaction instead of a constant potential does not qualitatively affect the thermopower result. We have also verified using the screened-Coulomb exact-diagonalization collision matrices of Ref.~\cite{nilsson25} that including higher radial modes does not change our results.

\paragraph{Matrix continued fraction.---} Sector by sector, the projected kinetic equation~\eqref{eq:projectedkinetic} reads
\begin{align}
L_1 {\bf h}_1 + \frac{i v_F q}{2} {\cal V} {\bf h}_{2}  &= - \frac{v_F}{T}
\bigl(e E_\perp {\bf c} + {\nabla}_\perp T  {\bf s} \bigr) \\
L_m {\bf h}_m + \frac{i v_F q}{2} {\cal V} {\bf h}_{m+1} &= - \frac{i v_F q}{2} {\cal V} {\bf h}_{m-1} .
\end{align}
Now assume that higher harmonics are sourced by the next-lower harmonics through the relation
\begin{align}
{\bf h}_{m+1} &= - {\cal G}_{m+1} \Bigl(\frac{i v_F q}{2} {\cal V}\Bigr) {\bf h}_m . \label{eq:defhm}
\end{align}
Inserting this into the above equation gives
\begin{align}
L_m {\bf h}_m + \Bigl(\frac{v_F q}{2}\Bigr)^2 {\cal V} {\cal G}_{m+1} {\cal V} {\bf h}_m &= - \Bigl(\frac{i v_F q}{2} {\cal V}\Bigr) {\bf h}_{m-1} .
\end{align}
Using now Eq.~\eqref{eq:defhm} for ${\bf h}_m$ and comparing the matrices gives Eq.~\eqref{eq:recursion}. For the endpoint $m=1$, comparison with the first equation above gives
\begin{align}
L_1 {\bf h}_1 + \Bigl(\frac{v_F q}{2}\Bigr)^2 {\cal V} {\cal G}_2 {\cal V} {\bf h}_1  &= - \frac{v_F}{T}
\bigl(e E_\perp {\bf c} + {\nabla}_\perp T  {\bf s} \bigr) .
\end{align}

\begin{figure}[t]
	\centering
    \scalebox{0.95}{\includegraphics{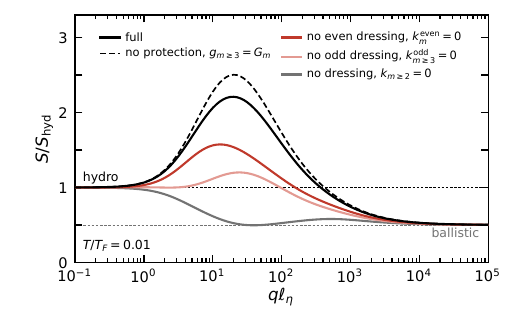}}
    \caption{Thermopower for $T/T_F=0.01$ with modified collision parameters [Eq.~\eqref{eq:collisionmatrix}]. Black: full result. Dashed black: Slow odd-mode rates set equal to the fast rates, $g_m=G_m$ for odd $m\geq 3$. Red lines: $k_m=0$ for even $m$ and for odd $m\geq 3$. Gray: $k_m=0$ for all $m\geq 2$.
    }
    \label{fig:4}
\end{figure}

\paragraph{Solving the matrix continued fraction.---} The chain for ${\cal G}_m$ is solved by assuming that for $m \geq m_c$, all odd-$m$ collision matrices are equal, as are all even ones, which gives
\begin{align}
{\cal G}_{\rm odd} &= \biggl[L_{\rm odd} + \Bigl(\frac{v_F q}{2}\Bigr)^2 {\cal V} {\cal G}_{\rm even} {\cal V}\biggr]^{-1} , \\
{\cal G}_{\rm even} &= \biggl[L_{\rm even} + \Bigl(\frac{v_F q}{2}\Bigr)^2 {\cal V} {\cal G}_{\rm odd} {\cal V}\biggr]^{-1} .
\end{align}
If we furthermore assume that $L$ is proportional to the identity, with $L_{\rm odd} = \gamma_o I$ and $L_{\rm even} = \gamma_e I$, the equations can be solved in closed form. We diagonalize the advection matrix
\begin{align}
U^T {\cal V} U &= \Lambda = \begin{pmatrix} \lambda_1 & 0 \\ 0 & \lambda_2 \end{pmatrix} ,
\end{align}
where $\lambda_1 = 1 + \kappa$ and $\lambda_2 = 1 - \kappa$ and
\begin{align}
U &= \frac{1}{\sqrt{2}} \begin{pmatrix} 1 & 1 \\ 1 & -1 \end{pmatrix} .
\end{align}
With $\tilde{\cal G}_{\rm odd} = U^T {\cal G}_{\rm odd} U$ and likewise $\tilde{\cal G}_{\rm even}$, we obtain
\begin{align}
\tilde{\cal G}_{\rm odd} &= \biggl[\gamma_{o} I + \Bigl(\frac{v_F q}{2}\Bigr)^2 \Lambda \tilde{\cal G}_{\rm even} \Lambda \biggr]^{-1} , 
\end{align}
which are solved by diagonal matrices $\tilde{\cal G}_{\rm odd} = {\rm diag} \{g_o(\lambda_1), g_o(\lambda_2)\}$ 
with
\begin{align}
g_o(\lambda) &= \biggl[\gamma_o + \lambda^2 \Bigl(\frac{v_F q}{2}\Bigr)^2 g_e(\lambda)\biggr]^{-1} .
\end{align}
These are quadratic equations for $g_e$ and $g_o$ with the solution
\begin{align}
g_o(\lambda) &
= \frac{2 \gamma_e}{\gamma_e \gamma_o \pm \sqrt{(\gamma_e \gamma_o)^2 + 4 \gamma_e \gamma_o \bigl(\lambda^2 \frac{v_F^2q^2}{4}\bigr)}} ,
\end{align}
where we need to pick the upper sign to ensure that in the limit $q\to 0$, we recover $g_e \simeq 1/\gamma_e$ and $g_o \simeq 1/\gamma_o$. The matrices are then iterated downward from $m_c$ to $m=1$, where we use as starting index $m_c=200$ with the mean of the diagonal elements of $L_{m_c}$ for $\gamma_e$ and $\gamma_o$. Results do not depend sensitively on this choice.

\begin{figure}
	\centering
    \scalebox{0.95}{\includegraphics{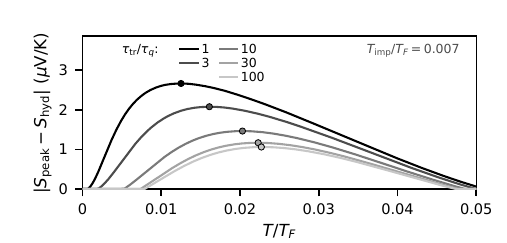}}
    \caption{Excess thermopower as a function of temperature at a fixed transport rate $T_{\rm imp}/T_F=0.007$, for different ratios of the quantum and transport lifetimes. Dots indicate the maxima. $\tau_{\rm tr}/\tau_{\rm q} = 1$ (black line) corresponds to short-range disorder, as in Fig.~\ref{fig:3}.}
    \label{fig:5}
\end{figure}

\paragraph{Origin of the thermopower overshoot.---} 
To isolate the origin of the thermopower overshoot numerically, we selectively alter the coefficients in the collision matrix~\eqref{eq:collisionmatrix}. Figure~\ref{fig:4} shows the same thermopower plot as in Fig.~\ref{fig:2}(b) for $T/T_F=0.01$ as a black line. To remove the effect of the collinear protection of odd modes, we set $g_m = G_m$ for odd $m \geq 3$ (black dashed line). The thermopower peak is enhanced compared to the full model, so the long odd-mode lifetimes reduce the thermopower overshoot. For the two red lines, we set $k_m=0$ for even $m$ and for odd $m\geq 3$, respectively. Here, we see a strong reduction in the thermopower peak. Even- and odd-mode dressings are additive (relative to the undressed response), with odd modes responsible for the larger part of the peak. If the dressing is removed for all $m\neq 1$ [gray line in Fig.~\ref{fig:4}], the peak is entirely suppressed and the thermopower crosses over from the hydrodynamic to the ballistic value without an overshoot. This is consistent with a recent study of the nonlocal thermopower within a Callaway model~\cite{beardo26}, which includes momentum-relaxing and momentum-conserving rates but no thermal dressing.

\paragraph{Angle-dependent disorder.---} A more general disorder model is $\gamma_m^{\rm dis} = \gamma_{\rm q} (1-\exp[- \varphi_0^2 m^2/2])$, where $\gamma_{\rm q} = 1/\tau_{\rm q}$ is the quantum relaxation rate and $\varphi_0$ is fixed by $\gamma_1^{\rm dis}=\gamma^{\rm dis}=1/\tau_{\rm tr}$, such that $\tau_{\rm tr}/\tau_{\rm q} = 1$ recovers short-range disorder.  Figure~\ref{fig:5} shows the excess thermopower for $\tau_{\rm tr}/\tau_{\rm q} = 1,3,10,30$, and $100$. 

\end{document}